\documentclass[aps,prd,twocolumn,nofootinbib,showpacs]{revtex4-1}

\usepackage{graphicx}

\usepackage[colorlinks=true,linkcolor=blue,citecolor=blue, urlcolor=blue]{hyperref}

\usepackage{amsmath}
\usepackage{multirow}
\usepackage{color}

\newcommand{\ree}{R_{e^+e^-}}
\newcommand{\rtau}{R_{\tau}}
\newcommand{\hbb}{\Gamma(H\to b\bar{b})}

\begin{document}

\title{Extending the Predictive Power of Perturbative QCD}

\author{Bo-Lun Du}
\email{dblcqu@cqu.edu.cn}
\author{Xing-Gang Wu}
\email{wuxg@cqu.edu.cn}
\author{Jian-Ming Shen}
\email{cqusjm@cqu.edu.cn}
\affiliation{Department of Physics, Chongqing University, Chongqing 401331, P.R. China}
\author{Stanley J. Brodsky}
\email{sjbth@slac.stanford.edu}
\affiliation{SLAC National Accelerator Laboratory, Stanford University, Stanford, California 94039, USA}

\begin{abstract}

The predictive power of perturbative QCD (pQCD) depends on two important issues: (1) how to eliminate the renormalization scheme-and-scale ambiguities at fixed order, and (2) how to reliably estimate the contributions of unknown higher-order terms using information from the known pQCD series. The Principle of Maximum Conformality (PMC) satisfies all of the principles of the renormalization group and eliminates the scheme-and-scale ambiguities by the recursive use of the renormalization group equation to determine the scale of the QCD running coupling $\alpha_s$ at each order. Moreover, the resulting PMC predictions are independent of the choice of the renormalization scheme, satisfying the key principle of renormalization group invariance. In this paper, we show that by using the conformal series derived using the PMC single-scale procedure, in combination with the Pad\'e Approximation Approach (PAA), one can achieve quantitatively useful estimates for the unknown higher-order terms from the known perturbative series. We illustrate this procedure for three hadronic observables $R_{e^+e^-}$, $R_{\tau}$, and $\Gamma(H \to b \bar{b})$ which are each known to 4 loops in pQCD. We show that if the PMC prediction for the conformal series for an observable (of leading order $\alpha_s^p$)  has been determined at order $\alpha^n_s$,  then the $[N/M]=[0/n-p]$ Pad\'e series provides quantitatively useful  predictions for the higher-order terms. We also show that the PMC + PAA predictions agree at all orders with the fundamental, scheme-independent Generalized Crewther relations which connect observables, such as  deep inelastic neutrino-nucleon scattering, to hadronic $e^+e^-$ annihilation. Thus, by using the combination of the PMC series and the Pad\'e method, the predictive power of pQCD theory can be greatly improved.



\end{abstract}

\maketitle

\subsection{Introduction}

Quantum chromodynamics (QCD) is believed to be the fundamental field theory of the hadronic strong interactions. Due to asymptotic freedom~\cite{Gross:1973id, Politzer:1973fx}, the QCD running coupling becomes numerically small at short distances, allowing perturbative calculations of observables for physical processes at large momentum transfer. The fundamental principle of renormalization group invariance requires that the prediction for a physical observable must be independent of both the choice of renormalization scheme and the choice of initial renormalization scale. However, due to the mismatch of the QCD running coupling ($\alpha_s$) and the pQCD coefficients at each order, a truncated pQCD series will not automatically satisfy this requirement, leading to well-known ambiguities. The predictive power of pQCD theory thus depends heavily on how to eliminate both the renormalization scheme-and-scale ambiguities and how to predict contributions from unknown higher-order terms.

It has become conventional to choose the renormalization scale $\mu_r$ as the typical momentum flow of the process. The resulting prediction at any fixed order will then inevitably also depend on the choice of the renormalization scheme. The hope is to achieve a nearly scheme-and-scale independent prediction by systematically computing higher-and-higher order QCD corrections; however, this hope is in direct conflict with the presence of the divergent  $n! \alpha_s^n \beta_0^n$ ``renormalon" series~\cite{Beneke:1994qe, Neubert:1994vb, Beneke:1998ui}. It is also often argued that by varying the renormalization scale, one will obtain information on the uncalculated higher-order terms. However, the variation of the renormalization scale can only provide information on the $\beta$-dependent terms which control the running of $\alpha_s$; the variation of $\mu_r$ gives no information on the contribution to the observable coming from the $\beta$-independent terms. We will refer to the $\beta$-independent contributions as ``conformal"  terms -- since they match the contributions of a corresponding conformal theory with $\beta=0$.

Obviously, the naive procedure of guessing and varying the renormalization scale can lead to a misleading pQCD prediction, especially if the conformal terms in the higher-order series are more important than the $\beta$-dependent terms. For example, the large $K$-factors for certain processes are caused by large conformal contributions,  as observed in the recent analysis of the $\gamma\gamma^*\to\eta_c$ transition form factor~\cite{Wang:2018lry}. Even if  a nearly scale-independent prediction is attained for a global quantity such as a total cross-section or a total decay width, the scale independence could be due to accidental cancellations among different orders, even though the scale dependence at each order could be very large. Worse, even if a prediction with a guessed scale agrees with the data, one cannot explain why it is reliable prediction, thus greatly depressing the predictive power of pQCD.

\subsection{The PMC}

The ``Principle of Maximum Conformality" (PMC)  rigorously eliminates the conventional renormalization scheme-and-scale ambiguities~\cite{Brodsky:2011ta, Brodsky:2012rj, Mojaza:2012mf, Brodsky:2013vpa}. It extends the well-known Brodsky-Lepage-Mackenzie (BLM) scale-setting method~\cite{Brodsky:1982gc} to all orders in pQCD. The basic PMC procedure is to identify all contributions which originate from the $\{\beta_i\}$-terms in a pQCD series; one then shifts the scale of the QCD running coupling at each order to absorb the $\{\beta_i\}$-terms and to thus obtain the correct scale for its running behavior as well as to set the number of active quark flavors $n_f$ arising from quark loops in the gluon propagators. The PMC also agrees with the standard Gell-Mann-Low method (GM-L)~\cite{GellMann:1954fq} for fixing the renormalization scale of $\alpha(Q^2)$ and the effective number of lepton flavors $n_\ell$ in Abelian quantum electrodynamics (QED).

One can choose any value for the initial renormalization scale $\mu_r$ when applying the PMC: the resulting scales for the running QCD coupling at each order are in practice independent of its value; thus the PMC eliminates the renormalization scale ambiguity. Moreover, the PMC predictions  are scheme-independent due to its  conformal nature, and the divergent renormalon behavior of the resulting perturbative series does not appear.

The PMC satisfies renormalization group invariance and all of the self-consistency conditions of the renormalization group equation (RGE)~\cite{Brodsky:2012ms}. The transition scale between the perturbative and nonperturbative domains can also be determined by using the PMC~\cite{Deur:2014qfa, Deur:2016cxb, Deur:2017cvd}, thus providing a physical procedure for setting the factorization scale for pQCD evolution. The PMC has now been successfully applied to many QCD measurements studies at the LHC as well as other hadronic  processes~\cite{Wu:2013ei, Wu:2014iba, Wu:2015rga}.

Within the framework of the PMC, the pQCD approximant can be written in the following form~\cite{Brodsky:2013vpa, Mojaza:2012mf},
\begin{eqnarray}
\rho_{n}(Q)|_{\rm Conv.} &=& \sum^{n}_{i=1} r_i(\mu^2_r/Q^2) a^{p+i-1}(\mu_r)   \label{rho}  \\
&=& r_{1,0}{a^p(\mu_r)} + \left[ r_{2,0} + p \beta_0 r_{2,1} \right]{a^{p+1}(\mu_r)} \nonumber\\
&& + \big[ r_{3,0} + p \beta_1 r_{2,1} + (p+1){\beta _0}r_{3,1} +\nonumber\\
&&  \frac{p(p+1)}{2} \beta_0^2 r_{3,2} \big]{a^{p+2}(\mu_r)} + \big[ r_{4,0} + p{\beta_2}{r_{2,1}}  \nonumber   \\
&& + (p+1){\beta_1}{r_{3,1}} + \frac{p(3+2p)}{2}{\beta_1}{\beta_0}{r_{3,2}} \nonumber\\
&& + (p+2){\beta_0}{r_{4,1}}+ \frac{(p+1)(p+2)}{2}\beta_0^2{r_{4,2}} \nonumber   \\
&& + \frac{p(p+1)(p+2)}{3!}\beta_0^3{r_{4,3}} \big]{a^{p+3}(\mu_r)} + \cdots,
\label{degeneracyrho}
\end{eqnarray}
where $a={\alpha_s}/\pi$, and $Q$ represents the kinematic scale. The index $p(\ge1)$ indicates the $\alpha_s$-order of the leading-order (LO) contribution and $\{r_{i,0}\}$ are conformal coefficients, and the $\beta$-pattern at each order is predicted by the non-Abelian gauge theory~\cite{Bi:2015wea}.

Following the standard PMC-s procedure, we obtain
\begin{eqnarray}
\rho_n(Q)|_{\rm PMC}=\sum_{i=1}^n r_{i,0}a^{p+i-1}(Q_{*}),
\label{pmcs}
\end{eqnarray}
where $Q_{*}$ is the determined optimal single PMC scale, whose analytical form can be found in Ref.~\cite{Shen:2017pdu}. We emphasize that the factorially divergent renormalon terms, such as $ n!\alpha_s^n \beta_0^n$, do not appear in the resulting conformal series; thus a convergent pQCD series can be achieved~\footnote{Only those $\{\beta_i\}$-terms that are pertained to RGE have been absorbed into the PMC scale. There may have cases in which the $\{\beta_i\}$-terms are not pertained to RGE and should be treated as conformal coefficients, which may break the pQCD convergence.}.

\subsection{Pad\'e Resummation}

The Pad\'e approximation approach (PAA)  provides a systematic procedure for promoting a finite Taylor series to an analytic function~\cite{Basdevant:1972fe, Samuel:1992qg, Samuel:1995jc}. In particular, the PAA can be used to estimate the $(n+1)_{\rm th}$-order coefficient by incorporating all known coefficients up to order $n$. It was shown in Ref.~\cite{Brodsky:1997vq} that the Pad\'e method provides an important guide for understanding the sequence of renormalization scales determined by the BLM method and its all-order extension, the PMC. Those scales are the optimal ones for evaluating each term in a skeleton expansion. The leading-order BLM/PMC sequence corresponds to the $[0/1]$-type PAA~\cite{Gardi:1996iq}. After applying the BLM/PMC, the summation over skeleton graphs is then similar to the summation of  the perturbative contributions for a corresponding conformal theory.

Since the divergent renormalon series does not appear in the conformal $\beta=0$ perturbative series generated by the PMC, there is an opportunity to use a resummation procedure such as the Pad\'e method to predict higher-order terms and to thus increase the precision and reliability of pQCD predictions. In this paper we will test whether one can use the PAA to achieve reliable predictions for the unknown higher-order terms for a pQCD series by using the renormalon-free conformal series determined by the PMC. For this purpose, we will adopt the PMC single-scale approach (PMC-s)~\cite{Shen:2017pdu}, which utilizes a single effective renormalization scale which matches the PMC series via the mean-value theorem.

Other applications of resummation methods to pQCD, together with alternatives to the PAA, have been discussed in the literature~\cite{Brodsky:1997vq, Gardi:1996iq, Ellis:1997sb, Burrows:1996dk, Ellis:1996zn, Jack:1997jn, Boito:2018rwt}. However, in our analysis, we will apply the PAA to the scale- and scheme- independent conformal series, whose perturbative coefficients are free of divergent renormalon contributions.

\subsection{Applying the PAA to pQCD}

If we apply the PAA to the PMC prediction, the pQCD series can be rewritten in the following $[N/M]$-type form
\begin{eqnarray}
\rho^{[N/M]}_n(Q)  &=&  a^p \times \frac{b_0+b_1 a + \cdots + b_N a^N}{1 + c_1 a + \cdots + c_M a^M}   \label{PAAseries0} \\
                               &=& \sum_{i=1}^{n} C_{i} a^{p+i-1} + C_{n+1}\; a^{p+n}+\cdots,  \label{PAAseries}
\end{eqnarray}
where $M\geq 1$ and $N+M+1 = n$. Comparing Eq.~(\ref{PAAseries}) with the series (\ref{rho}) or (\ref{pmcs}), the coefficients $C_{i}$ can be directly related to $r_i$ or $r_{i,0}$, respectively. Furthermore, by using the known ${\rm N^{n-1}LO}$-order pQCD series, the coefficients $b_{i\in[0,N]}$ and $c_{i\in[1,M]}$ can be expressed by using the coefficients $C_{i\in[1,n]}$. Finally, we can use the coefficients $b_{i\in[0,N]}$ and $c_{i\in[1,M]}$ to predict the one-order-higher uncalculated coefficient $C_{n+1}$ at the ${\rm N^{n}LO}$-order level. For examples, if $[N/M]=[n-2/1]$, we have
\begin{equation}
C_{n+1}=\frac{C_n^2}{C_{n-1}}; \label{n-2/1}
\end{equation}
if $[N/M]=[n-3/2]$, we have
\begin{eqnarray}
C_{n+1}=\frac{-C_{n-1}^3+2C_{n-2}C_{n-1}C_{n}-C_{n-3}C_{n}^2}{C_{n-2}^2-C_{n-3}C_{n-1}}; \label{n-3/2}
\end{eqnarray}
if  $[N/M]=[n-4/3]$, we have
\begin{eqnarray}
&&C_{n+1}=\{C_{n-2}^4-(3 C_{n-3} C_{n-1}+2 C_{n-4} C_{n}) C_{n-2}^2 \nonumber \\
&&\quad\quad +2 [C_{n-4} C_{n-1}^2+(C_{n-3}^2+C_{n-5} C_{n-1}) C_{n}] C_{n-2} \nonumber \\
&&\quad\quad -C_{n-5} C_{n-1}^3+C_{n-3}^2 C_{n-1}^2+C_{n-4}^2 C_{n}^2 \nonumber \\
&&\quad\quad -C_{n-3} C_{n} (2 C_{n-4} C_{n-1}+C_{n-5} C_{n})\} \nonumber \\
&&\quad\quad / \{C_{n-3}^3-\left(2 C_{n-4} C_{n-2}+C_{n-5} C_{n-1}\right) C_{n-3} \nonumber \\
&&\quad\quad +C_{n-5} C_{n-2}^2+C_{n-4}^2 C_{n-1}\}; \label{n-4/3} {\rm etc.}
\end{eqnarray}
In each case, $C_{i<1}\equiv 0$. We need to know at least two $C_i$ in order to predict the unknown higher-order coefficients; thus the PAA is applicable when we have determined at least the NLO terms ($n=2$) using the PMC. One can also use the full PAA (\ref{PAAseries0}) to estimate the sum of the whole series, e.g. to give the all-oders PAA prediction. As will be found later, the differences for the predictions of the truncated and full PAA series are small for converged perturbative series.

In the following, we will apply the PAA for three physical observables $\ree$, $\rtau$ and $\hbb$ which are known at four loops in pQCD. We will show how the ``unknown" terms predicted by the PAA varies when one inputs more-and-more known higher-order terms.

The ratio $\ree$ is defined as
\begin{eqnarray}
R_{e^+ e^-}(Q) &=& \frac{\sigma\left(e^+e^-\to {\rm hadrons} \right)}{\sigma\left(e^+e^-\to \mu^+ \mu^-\right)}\nonumber\\
&=& 3\sum_q e_q^2\left[1+R(Q)\right], \label{ree}
\end{eqnarray}
where $Q=\sqrt{s}$ is the $e^+e^-$ collision energy. The pQCD approximants for $R(Q)$ are labelled $R_n(Q)= \sum_{i=1}^{n} r_i(\mu_r/Q)a^{i}(\mu_r)$. The pQCD coefficients at $\mu_r=Q$ have been calculated in the $\overline{\rm MS}$-scheme in Refs.~\cite{Baikov:2008jh, Baikov:2010je, Baikov:2012zm, Baikov:2012zn}. For illustration we take $Q=31.6 \;{\rm GeV}$~\cite{Marshall:1988ri}.

The ratio $\rtau$ is defined as
\begin{eqnarray}
R_{\tau}(M_{\tau}) &=&\frac{\sigma(\tau\rightarrow\nu_{\tau}+\rm{hadrons)}}{\sigma(\tau\rightarrow\nu_{\tau}+\bar{\nu}_e+e^-)}\nonumber\\
&=&3\sum\left|V_{ff'}\right|^2\left(1+\tilde{R}(M_{\tau})\right),
\end{eqnarray}
where $V_{ff'}$ are Cabbibo-Kobayashi-Maskawa matrix elements, $\sum\left|V_{ff'}\right|^2 =\left(\left|V_{ud}\right|^2+\left|V_{us}\right|^2\right)\approx 1$ and $M_{\tau}= 1.78$ GeV. The pQCD approximant, $\tilde{R}_{n}(M_{\tau})= \sum_{i=1}^{n}r_i(\mu_r/M_{\tau})a^{i}(\mu_r)$; the coefficients can be obtained by using the known relation of $R_{\tau}(M_{\tau})$ to $R(\sqrt{s})$~\cite{Lam:1977cu}.

The decay width $\hbb$ is defined as
\begin{eqnarray}
\Gamma(H\to b\bar{b})=\frac{3G_{F} M_{H} m_{b}^{2}(M_{H})} {4\sqrt{2}\pi} [1+\hat{R}(M_{H})],   \label{hbb}
\end{eqnarray}
where the Fermi constant $G_{F}=1.16638\times10^{-5}\;\rm{GeV}^{-2}$, the Higgs mass $M_H=126$ GeV, and the $b$-quark $\overline{\rm{MS}}$-running mass is $m_b(M_H)=2.78$ GeV~\cite{Wang:2013bla}. The pQCD approximant $\hat{R}_n(M_H)= \sum_{i=1}^{n}r_i(\mu_r/M_{H}) a^{i}(\mu_r)$, where the predictions for the $\overline{\rm MS}$-coefficients at $\mu_r=M_H$ can be found in Ref.~\cite{Baikov:2005rw}.

In each case the coefficients at any other scale can be obtained via QCD evolution. In doing the numerical evaluation, we have assumed the running of $\alpha_s$   at the four-loop level. The asymptotic QCD scale is set using $\alpha_s(M_z)=0.1181$~\cite{Olive:2016xmw}, giving $\Lambda_{\rm{QCD}}^{n_f=5}=0.210$ GeV.

After applying the PMC-s approach, the optimal scale for each process can be determined. If the pQCD approximants are known at up to two-loop, three-loop, and four-loop level, the corresponding optimal scales are $Q_{*}|_{e^+e^-}=[35.36$, $39.68$, $40.30]$ GeV, $Q_{*}|_{\tau}=[0.90$, $1.01$, $1.05]$ GeV~\footnote{Because the usually adopted analytic $\alpha_s$-running differ significantly at scales below a few GeV from the exact solution of RGE~\cite{Olive:2016xmw}, we will use the exact numerical solution of the RGE throughout to evaluate $\rtau$.} and $Q_{*}|_{H\to b\bar{b}}=[61.38$, $57.41$, $58.84]$ GeV, accordingly. It is found that those PMC scales $Q^*$ are completely independent of  the choice of the initial renormalization scale $\mu_r$.

\begin{table}[htb]
\centering
\begin{tabular}{  |c| c| c| c |}
\hline
~$r_{n+1,0}$ ~&~~~$n+1=3$~~~ & ~~~$n+1=4$~~~ & ~~~$n+1=5$~~~\\
 \hline
~EC~  & $-1.0 $ & $-11.0 $ & - \\
\hline
~\multirow{3}{*}{PAA}~  & [0/1]$+3.4$ & [0/2]$-9.9$ &[0/3]$-17.8$ \\
\cline{2-4}
- & - &[1/1]$+0.55$ & [1/2]$-18.0$ \\
\cline{2-4}
  & - & - & [2/1]$-120.$ \\
\hline
\end{tabular}
\caption{Comparison of the exact (``EC") $(n+1)_{\rm th}$-order conformal coefficients with the predicted (``$[N/M]$-type PAA") $(n+1)_{\rm th}$-order ones based on the known $n_{\rm th}$-order approximate $R_{n}(Q=31.6~ {\rm GeV})$, where $n=2,3,4$, respectively.}
\label{estimate-ree}
\end{table}

\begin{table}[htb]
\centering
\begin{tabular}{ |c| c| c| c |}
\hline
~$r_{n+1,0}$ ~&~~~$n+1=3$~~~ & ~~~$n+1=4$~~~ & ~~~$n+1=5$~~~\\
\hline
~EC~ & $+3.4$ & $+6.8$ &  - \\
\hline
~\multirow{3}{*}{PAA}~ & [0/1]$+4.6$ & [0/2]$+4.9$ &[0/3]$+14.7$  \\
\cline{2-4}
   &- &[1/1]$+5.5$ & [1/2]$+11.5$ \\
\cline{2-4}
& - & - &[2/1]$+13.5$ \\
\hline
\end{tabular}
\caption{Comparison of the exact (``EC") $(n+1)_{\rm th}$-order conformal coefficients with the predicted (``$[N/M]$-type PAA") $(n+1)_{\rm th}$-order ones based on the known $n_{\rm th}$-order approximate $\tilde{R}_{n}(M_{\tau})$, where $n=2,3,4$, respectively.}
\label{estimate-rtau}
\end{table}

\begin{table}[htb]
\centering
\begin{tabular}{|c| c| c| c |}
\hline
~$r_{n+1,0}$ ~&~~~$n+1=3$~~~ & ~~~$n+1=4$~~~ & ~~~$n+1=5$~~~\\
\hline
~EC~  & $-1.36\times10^2$ & $-4.32\times10^2$ &  -   \\
\hline
~\multirow{3}{*}{PAA}~  & [0/1]$+3.23\times10^1$ & [0/2]$-7.26\times10^2$ &[0/3]$+3.72\times10^3$ \\
\cline{2-4}
& - & [1/1]$+1.37\times10^3$ &[1/2]$+3.20\times10^3$ \\
\cline{2-4}
& -  & - &[2/1]$-1.37\times10^3$ \\
\hline
\end{tabular}
\caption{Comparison of the exact (``EC") $(n+1)_{\rm th}$-order conformal coefficients with the predicted (``$[N/M]$-type PAA") $(n+1)_{\rm th}$-order ones based on the known $n_{\rm th}$-order approximate $\hat{R}_{n}(M_H)$, where $n=2,3,4$, respectively. }
\label{estimate-hbb}
\end{table}

The remaining task for the PAA is to predict the higher-order conformal coefficients. We present a comparison of the exact $(n+1)_{\rm th}$-order conformal coefficients with the PAA predicted ones based on the known $n_{\rm th}$-order approximates $R_{n}(Q=31.6~ {\rm GeV})$, $\tilde{R}_{n}(M_{\tau})$ and $\hat{R}_{n}(M_H)$ in Tables~\ref{estimate-ree}, ~\ref{estimate-rtau} and~\ref{estimate-hbb}, respectively.  Here the $[N/M]$-type PAA is for $N+M=n-1$ with $N\geq0$ and $M\geq1$. Those Tables show that the $[N/M]=[0/n-1]$-type PAA provides result closest to the known pQCD result.

It is interesting to note that the $[0/n-1]$-type PAA is consistent with the ``Generalized Crewther Relations" (GSICRs)~\cite{Shen:2016dnq}. For example, the GSICR, which provides a remarkable all-orders connection between the pQCD predictions for deep inelastic neutrino-nucleon scattering and hadronic $e^+e^-$ annihilation shows that the conformal coefficients are all equal to $1$; e.g. $\widehat{\alpha}_d(Q)=\sum_{i}\widehat{\alpha}^{i}_{g_1}(Q_*)$, where $Q_*$ satisfies
\begin{eqnarray}
\ln\left.\frac{Q_*^2}{Q^2}\right|_{g_1} &=& 1.308 + [-0.802 + 0.039 n_f] \widehat{\alpha}_{g_1}(Q_*) + \nonumber\\
&&  [16.100 - 2.584 n_f + 0.102 n_f^2] \widehat{\alpha}_{g_1}^2(Q_*)  + \cdots.
\end{eqnarray}
By using the $[0/n-1]$-type PAA -- the geometric series -- all of the predicted conformal coefficients are also equal to $1$.

The $[0/n-1]$-type PAA also agrees with the GM-L scale-setting procedure to obtain scale-independent perturbative QED predictions; e.g., the renormalization scale for the electron-muon elastic scattering through one-photon exchange is set as the virtuality of the exchanged photon, $\mu_r^2 = q^2 = t$. By taking an arbitrary initial renormalization scale $t_0$, we have
\begin{equation}
\alpha_{em}(t) = \frac{\alpha_{em}(t_0)}{1 - \Pi(t,t_0)} \;\;,
\end{equation}
where $\Pi(t,t_0) = \frac{\Pi(t,0) -\Pi(t_0,0)}{1-\Pi(t_0,0)}$, which sums all vacuum polarization contributions, both proper and improper, to the dressed photon propagator. The PMC reduces in the $N_C \to 0$ Abelian limit to the GM-L method~\cite{Brodsky:1997jk} and the preferable $[0/n-1]$-type makes the PAA geometric series self-consistent with the GM-L/PMC prediction.

Tables~\ref{estimate-ree}, ~\ref{estimate-rtau} and~\ref{estimate-hbb} show that as more loop terms are inputted, the predicted conformal coefficients become closer to their exact value. To show this clearly, we define the normalized difference between the exact conformal coefficient and the predicted one as
\begin{displaymath}
\Delta_{n} = \left|\frac{r_{n,0}|_{\rm PAA}-r_{n,0}|_{\rm EC}}{r_{n,0}|_{\rm EC}}\right| ,
\end{displaymath}
where ``EC" and ``PAA" stand for exact and predicted conformal coefficients, respectively. By using the exact terms, known up to two-loop and three-loop levels accordingly, the normalized differences for the $3_{\rm th}$-order and the $4_{\rm th}$-order conformal coefficients, i.e. those coefficients in the $n+1=3$ and $n+1=4$ columns in Tables~\ref{estimate-ree}-\ref{estimate-hbb}, become suppressed from $440.\%$ to $10\%$ for $R(Q=31.6~{\rm GeV})$, from $35\%$ to $28\%$ for $\tilde{R}(M_{\tau})$, and from $124.\%$ to $68\%$ for $\hat{R}(M_H)$. There are large differences for the conformal coefficients if we only know the QCD corrections at the two-loop level; however this decreases rapidly when we know more loop terms. Following this trend, the normalized differences for the $5_{\rm th}$-order conformal coefficients should be much smaller than the $4_{\rm th}$-order ones. Conservatively, if we set the normalized difference ($\Delta_5$) of the $5_{\rm th}$-loop as the same one of the $4_{\rm}$-loop ($\Delta_4$), we can inversely predict the $5_{\rm th}$-loop ``$\rm{EC'}$" conformal coefficients:
\begin{eqnarray}
r^{e^+ e^-}_{5,0}|_{\rm EC'} &=& -18.0\pm1.8, \\
r^{\tau}_{5,0}|_{\rm EC'} &=& 16.0\pm 4.5, \\
r^{H\to b\bar{b}}_{5,0}|_{\rm EC'} &=& (6.92\pm4.71)\times10^3,
\end{eqnarray}
where the central values are obtained by averaging the two ``${\rm EC'}$" values determined by $\frac{r^{e^+ e^-}_{5,0}|_{\rm PAA}}{(1\pm\Delta_4)}$.

\begin{widetext}
\begin{center}
\begin{table}[htb]
\begin{tabular}{ccccccc}
\hline
 &~~${\rm EC}$, $n=2$~~&~~${\rm PAA}$, $n=3$~~&~~${\rm EC}$, $n=3$~~&~~${\rm PAA}$, $n=4$~~&~~ ${\rm EC}$, $n=4$~~&~~${\rm PAA}$, $n=5$~~ \\
 \hline
$R_n(Q)|_{\rm PMC-s}$ & 0.04745 & 0.04772(0.04777) & 0.04635 & 0.04631(0.04631) & 0.04619 & 0.04619(0.04619)  \\
$\tilde{R}_{n}(M_{\tau})|_{\rm PMC-s}$ & 0.1879  &  0.2035(0.2394)  &  0.2103  &  0.2128(0.2134)  & 0.2089   & 0.2100(0.2104)    \\
$\hat{R}_{n}(M_H)|_{\rm PMC-s}$  & 0.2482  &  0.2503(0.2505)  &  0.2422  &   0.2402(0.2406)  & 0.2401   & 0.2405(0.2405)    \\
$R_n(Q)|_{\rm Conv.}$  & $0.04763^{+0.00045}_{-0.00139}$ & $0.04781^{+0.00043}_{-0.00053}$ & $0.04648_{-0.00071}^{+0.00012}$ & $0.04632_{-0.00025}^{+0.00018}$ &$0.04617_{-0.00009}^{+0.00015}$ &$0.04617_{-0.00001}^{+0.00007}$ \\
$\tilde{R}_{n}(M_{\tau})|_{\rm Conv.}$  & $0.1527^{+0.0610}_{-0.0323}$ & $0.1800^{+0.0515}_{-0.0330}$ & $0.1832_{-0.0334}^{+0.0385}$ & $0.1975_{-0.0296}^{+0.0140}$ & $0.1988_{-0.0299}^{+0.0140}$ & $0.2056_{-0.0247}^{+0.0029}$ \\
$\hat{R}_{n}(M_H)|_{\rm Conv.}$   & $0.2406^{+0.0074}_{-0.0104}$ & $0.2475^{+0.0027}_{-0.0066}$ & $0.2425_{-0.0053}^{+0.0002}$ & $0.2419_{-0.0040}^{+0.0002}$ & $0.2411_{-0.0040}^{+0.0001}$ & $0.2407_{-0.0040}^{+0.0002}$ \\
\hline
\end{tabular}
\caption{Comparison of the exact (``EC") and the predicted (``PAA") pQCD approximants $R_n(Q=31.6\;{\rm GeV})$,  $\tilde{R}_{n}(M_\tau)$ and $\hat{R}_n(M_H)$ under conventional (Conv.) and PMC-s scale-setting approaches up to $n_{\rm th}$-order level. The $(n+1)_{\rm th}$-order PAA prediction equals to the $n_{\rm th}$-order known prediction plus the predicted $(n+1)_{\rm th}$-order terms using the $[0/n-1]$-type PAA prediction (The values in the parentheses are results for the corresponding full PAA series). The PMC predictions are scale independent and the errors for conventional scale-setting are estimated by varying the initial renormalization scale $\mu_r$ within the region of $[1/2\mu_0, 2\mu_0]$, where $\mu_0=Q$, $M_\tau$ and $M_H$, respectively. }
\label{Total-observable}
\end{table}
\end{center}
\end{widetext}

The difference between the exact and predicted conformal coefficients is reduced by the  $\alpha_s/\pi$-power suppression, thus the precision of the predictive power of the PAA should become most useful for total cross-sections and decay widths. We present the comparison of the exact results for $R_{n}(Q=31.6~{\rm GeV})$, $\tilde{R}_{n}(M_{\tau})$ and $\hat{R}_{n}(M_H)$ with the $[0/n-1]$-type PAA predicted ones in Table~\ref{Total-observable}.  The values in the parentheses are results for the corresponding full PAA series, which are calculated by using Eq.~(\ref{PAAseries0}). Due to the fast pQCD convergence, the differences between the truncated and full PAA predictions are small, which are less than $1\%$ for $n\ge 4$. Similarly, we define the precision of the predictive power as the normalized difference between the exact approximant ($\rho_n|_{\rm{EC}}$) and the prediction ($\rho_n|_{\rm{PAA}}$); i.e.
\begin{displaymath}
\left|\frac{\rho_{n}|_{\rm PAA}- \rho_{n}|_{\rm EC}}{\rho_{n}|_{\rm EC}}\right|.
\end{displaymath}

The PMC predictions are renormalization scheme-and-scale independent, and the pQCD convergence is greatly improved due to the elimination of renormalon contributions. Highly precise values at each order can thus be achieved~\cite{Shen:2017pdu}. In contrast,  predictions using conventional pQCD series (\ref{rho}) are scale dependent even for higher-order predictions. We also present results using conventional scale-setting in Table~\ref{Total-observable}; it  confirms the conclusion that the conformal PMC-s series is much more suitable for applications of the PAA.

By using the known (exact) approximants predicted by PMC-s scale-setting up to two-loop and three-loop levels accordingly, the differences between the exact and predicted three-loop and four-loop approximants are observed to decrease from $3.0\%$ to $0.3\%$ for $\rho_n=R_n(Q=31.6~{\rm GeV})$, from $3\%$ to $2\%$ for $\rho_n=\tilde{R}_n(M_{\tau})$, and from $3.0\%$ to $\sim 0\%$ for $\rho_n=\hat{R}_n(M_H)$, respectively.  The normalized differences for $R_4(Q=31.6~{\rm GeV})$, $\tilde{R}_4(M_{\tau})$ and $\hat{R}_4(M_H)$ are small. If we conservatively set the normalized difference of the $5_{\rm th}$-loop to match that of the $4_{\rm}$-loop predictions, then the predicted $5_{\rm th}$-loop ``$\rm{EC'}$" predictions are
\begin{eqnarray}
R_5(Q=31.6~{\rm GeV})|_{\rm EC'} &=& 0.04619\pm0.00014, \label{five-ee}\\
\tilde{R}_5(M_{\tau})|_{\rm EC'} &=& 0.2100\pm0.0042, \\
\hat{R}_5(M_H)|_{\rm EC'} &=& 0.2405\pm0.0001.
\end{eqnarray}

\subsection{Summary}

The PMC provides first-principle predictions for QCD; it satisfies renormalization group invariance and eliminates the conventional renormalization scheme-and-scale ambiguities. Since the divergent renormalon series does not appear in the conformal ($\beta=0$) perturbative series generated by the PMC, there is an opportunity to use resummation procedures such as the Pad\'e method to predict higher-order terms and thus to increase the precision and reliability of pQCD predictions.

In this paper, we have shown that by applying PAA to the renormalon-free conformal series derived by using the PMC single-scale procedure, one can achieve quantitatively useful estimates for the unknown higher-order terms based on the known perturbative QCD series.

In particular, we have found that if the PMC prediction for the conformal series for an observable (of leading order $\alpha_s^p$)  has been determined at order $\alpha^n_s$, then the $[N/M]=[0/n-p]$ Pad\'e series provides an important estimate for the higher-order terms. The all-orders predictions of the $[0/n-p]$-type PAA are in fact identical to the predictions obtained from the all-order GSICRs which connect observables, such as deep inelastic neutrino-nucleon scattering, to hadronic $e^+e^-$ annihilation. These relations are fundamental, high precision predictions of QCD.

\begin{figure}[htb]
\centering
\includegraphics[width=0.45\textwidth]{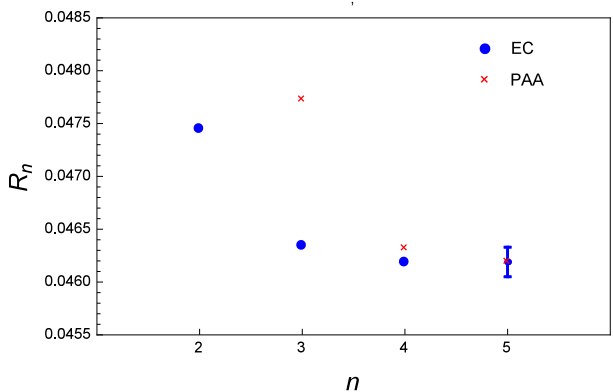}
\caption{Comparison of the exact (``EC") and the predicted ([0/n-1]-type ``PAA") pQCD prediction for $R_n(Q=31.6\;{\rm GeV})$ under the PMC-s scale-setting. It shows how the PAA predictions change when more loop-terms are included, where the five-loop ``EC" prediction is from Eq.~(\ref{five-ee}). } \label{Fig:Ree}
\end{figure}

Tables~\ref{estimate-ree}, ~\ref{estimate-rtau} and~\ref{estimate-hbb} show that the difference between the exact and the predicted conformal coefficients at  various loops, which decreases rapidly as additional high-order loop terms are included. Table~\ref{Total-observable} shows that the PAA becomes quantitatively effective even at the NLO level for the pQCD approximant due to the strong $\alpha_s/\pi$-suppression of the conformal series. For example, when using the NLO results $R_2(Q)$, $\tilde{R}_2(M_{\tau})$ and $\hat{R}_2(M_H)$ to predict the observables $R_3(Q)$, $\tilde{R}_3(M_{\tau})$ and $\hat{R}_3(M_H)$ at NNLO, the normalized differences between the Pad\'e estimates and the known results are only about $3\%$. Taking $\ree$ as an explicit example, we show how the PAA predictions change when more loop-terms are included in Fig.~\ref{Fig:Ree}. In some sense this is an infinite-order prediction for $R_{e^+e^-}(Q=31.6\;{\rm GeV})$, and it is the most precise prediction one can make using our PMC+PAA method, given the present knowledge of pQCD. Thus by combining the PMC with the Pad\'e method, the predictive power of the pQCD theory can be remarkably improved.

As a final remark, we show that the way of using PAA basing on the conformal series is consistent with that of the ${\cal N}=4$ supersymmetric Yang-Mills theory. For the purpose, we present a PAA prediction on the NNLO and $\rm{N^3LO}$ Balitsky-Fadin-Kuraev-Lipatov (BFKL) Pomeron eigenvalues. By using the PAA method together with the known LO and NLO coefficients given in Ref.~\cite{Costa:2012cb}, we find that the NNLO BFKL coefficient is $0.86\times10^4$ for $\Delta=0.45$, where $\Delta$ is the full conformal dimension of the twisted-two operator. The exact NNLO BFKL coefficient has been discussed in planar ${\cal N}=4$ supersymmetric Yang-Mills theory~\cite{Gromov:2015vua} by using the quantum spectral curve integrability-based method~\cite{Gromov:2013pga, Gromov:2014caa}, which gives $1.08\times10^4$~\cite{Gromov:2015vua}. Thus the normalized difference between those two NNLO values is only about $20\%$. As a step forward, we predict the $\rm{N^3LO}$ coefficient to the Pomeron eigenvalue by using the $[0/2]$-PAA type and the known NNLO coefficient given in Ref.~\cite{Gromov:2015vua}, which results in $-3.07\times10^5$. This value is also consistent with the ${\cal N}=4$ supersymmetric Yang-Mills prediction, since if we adopt the data-fitting prediction suggested in Ref.~\cite{Gromov:2015wca} to predict $\rm{N^3LO}$ coefficient, we shall obtain $-3.66\times10^5$. The normalized difference between those two $\rm{N^3LO}$ values is also only about $20\%$.

\hspace{1cm}

\noindent{\bf Acknowledgments}: This work is supported in part by the Natural Science Foundation of China under Grant No.11625520 and No.11847301, the Fundamental Research Funds for the Central Universities under the Grant No.2018CDPTCG0001/3, and the Department of Energy Contract No. DE-AC02-76SF00515. SLAC-PUB-17306.


\begin{thebibliography}{100}

\bibitem{Gross:1973id}
  D.~J.~Gross and F.~Wilczek,
  Phys.\ Rev.\ Lett.\  {\bf 30}, 1343 (1973).

\bibitem{Politzer:1973fx}
  H.~D.~Politzer,
  Phys.\ Rev.\ Lett.\  {\bf 30}, 1346 (1973).

\bibitem{Beneke:1994qe}
  M.~Beneke and V.~M.~Braun,
  Phys.\ Lett.\ B {\bf 348}, 513 (1995).

\bibitem{Neubert:1994vb}
  M.~Neubert,
  Phys.\ Rev.\ D {\bf 51}, 5924 (1995).

\bibitem{Beneke:1998ui}
  M.~Beneke,
  Phys.\ Rept.\  {\bf 317}, 1 (1999).

\bibitem{Wang:2018lry}
  S.~Q.~Wang, X.~G.~Wu, W.~L.~Sang and S.~J.~Brodsky,
  Phys.\ Rev.\ D {\bf 97}, 094034 (2018).

\bibitem{Brodsky:2011ta}
  S.~J.~Brodsky and X.~G.~Wu,
  Phys.\ Rev.\ D {\bf 85}, 034038 (2012).

\bibitem{Brodsky:2012rj}
  S.~J.~Brodsky and X.~G.~Wu,
  Phys.\ Rev.\ Lett.\  {\bf 109}, 042002 (2012).

\bibitem{Mojaza:2012mf}
  M.~Mojaza, S.~J.~Brodsky and X.~G.~Wu,
  Phys.\ Rev.\ Lett.\  {\bf 110}, 192001 (2013).

\bibitem{Brodsky:2013vpa}
  S.~J.~Brodsky, M.~Mojaza and X.~G.~Wu,
  Phys.\ Rev.\ D {\bf 89}, 014027 (2014).

\bibitem{Brodsky:1982gc}
 S.~J.~Brodsky, G.~P.~Lepage and P.~B.~Mackenzie,
 Phys.\ Rev.\ D {\bf 28}, 228 (1983).

\bibitem{GellMann:1954fq}
  M.~Gell-Mann and F.~E.~Low,
  Phys.\ Rev.\  {\bf 95}, 1300 (1954).

\bibitem{Brodsky:2012ms}
  S.~J.~Brodsky and X.~G.~Wu,
  Phys.\ Rev.\ D {\bf 86}, 054018 (2012).

\bibitem{Deur:2014qfa}
  A.~Deur, S.~J.~Brodsky and G.~F.~de Teramond,
  Phys.\ Lett.\ B {\bf 750}, 528 (2015).

\bibitem{Deur:2016cxb}
  A.~Deur, S.~J.~Brodsky and G.~F.~de Teramond,
  Phys.\ Lett.\ B {\bf 757}, 275 (2016).

\bibitem{Deur:2017cvd}
  A.~Deur, J.~M.~Shen, X.~G.~Wu, S.~J.~Brodsky and G.~F.~de Teramond,
  Phys.\ Lett.\ B {\bf 773}, 98 (2017).

\bibitem{Wu:2014iba}
  X.~G.~Wu, Y.~Ma, S.~Q.~Wang, H.~B.~Fu, H.~H.~Ma, S.~J.~Brodsky and M.~Mojaza,
  Rept.\ Prog.\ Phys.\  {\bf 78}, 126201 (2015).

\bibitem{Wu:2015rga}
  X.~G.~Wu, S.~Q.~Wang and S.~J.~Brodsky,
  Front.\ Phys. {\bf 11}, 1, 111201 (2016).

\bibitem{Wu:2013ei}
  X.~G.~Wu, S.~J.~Brodsky and M.~Mojaza,
  Prog.\ Part.\ Nucl.\ Phys.\  {\bf 72}, 44 (2013).

  \bibitem{Bi:2015wea}
  H.~Y.~Bi, X.~G.~Wu, Y.~Ma, H.~H.~Ma, S.~J.~Brodsky and M.~Mojaza,
  Phys.\ Lett.\ B {\bf 748}, 13 (2015).

  \bibitem{Shen:2017pdu}
  J.~M.~Shen, X.~G.~Wu, B.~L.~Du and S.~J.~Brodsky,
  Phys.\ Rev.\ D {\bf 95}, 094006 (2017).

\bibitem{Basdevant:1972fe}
  J.~L.~Basdevant,
  Fortsch.\ Phys.\ {\bf 20}, 283 (1972).

\bibitem{Samuel:1992qg}
  M.~A.~Samuel, G.~Li and E.~Steinfelds,
  Phys.\ Lett.\ B {\bf 323}, 188 (1994).

\bibitem{Samuel:1995jc}
  M.~A.~Samuel, J.~R.~Ellis and M.~Karliner,
  Phys.\ Rev.\ Lett.\ {\bf 74}, 4380 (1995).

\bibitem{Brodsky:1997vq}
  S.~J.~Brodsky, J.~R.~Ellis, E.~Gardi, M.~Karliner and M.~A.~Samuel,
  Phys.\ Rev.\ D {\bf 56}, 6980 (1997).

\bibitem{Gardi:1996iq}
  E.~Gardi,
  Phys.\ Rev.\ D {\bf 56}, 68 (1997).

\bibitem{Ellis:1997sb}
  J.~R.~Ellis, I.~Jack, D.~R.~T.~Jones, M.~Karliner and M.~A.~Samuel,
  Phys.\ Rev.\ D {\bf 57}, 2665 (1998).

\bibitem{Burrows:1996dk}
  P.~N.~Burrows, T.~Abraha, M.~Samuel, E.~Steinfelds and H.~Masuda,
  Phys.\ Lett.\ B {\bf 392}, 223 (1997).

\bibitem{Ellis:1996zn}
  J.~R.~Ellis, M.~Karliner and M.~A.~Samuel,
  Phys.\ Lett.\ B {\bf 400}, 176 (1997).

\bibitem{Jack:1997jn}
  I.~Jack, D.~R.~T.~Jones and M.~A.~Samuel,
  Phys.\ Lett.\ B {\bf 407}, 143 (1997).

\bibitem{Boito:2018rwt}
  D.~Boito, P.~Masjuan and F.~Oliani,
  JHEP {\bf 1808}, 075 (2018).

\bibitem{Baikov:2008jh}
  P.~A.~Baikov, K.~G.~Chetyrkin and J.~H.~Kuhn,
  Phys.\ Rev.\ Lett.\ {\bf 101}, 012002 (2008).

\bibitem{Baikov:2010je}
  P.~A.~Baikov, K.~G.~Chetyrkin and J.~H.~Kuhn,
  Phys.\ Rev.\ Lett.\ {\bf 104}, 132004 (2010).

\bibitem{Baikov:2012zn}
  P.~A.~Baikov, K.~G.~Chetyrkin, J.~H.~Kuhn and J.~Rittinger,
  Phys.\ Lett.\ B {\bf 714}, 62 (2012).

\bibitem{Baikov:2012zm}
  P.~A.~Baikov, K.~G.~Chetyrkin, J.~H.~Kuhn and J.~Rittinger,
  JHEP {\bf 1207}, 017 (2012).

\bibitem{Marshall:1988ri}
  R.~Marshall,
  Z.\ Phys.\ C {\bf 43}, 595 (1989).

\bibitem{Lam:1977cu}
  C.~S.~Lam and T.~-M.~Yan,
  Phys.\ Rev.\ D {\bf 16}, 703 (1977).

\bibitem{Wang:2013bla}
  S.~Q.~Wang, X.~G.~Wu, X.~C.~Zheng, J.~M.~Shen and Q.~L.~Zhang,
  Eur.\ Phys.\ J.\ C {\bf 74}, 2825 (2014).

\bibitem{Baikov:2005rw}
  P.~A.~Baikov, K.~G.~Chetyrkin and J.~H.~Kuhn,
  Phys.\ Rev.\ Lett.\ {\bf 96}, 012003 (2006).

\bibitem{Olive:2016xmw}
  C.~Patrignani {\it et al.} [Particle Data Group],
  Chin.\ Phys.\ C {\bf 40}, 100001 (2016).

\bibitem{Shen:2016dnq}
  J.~M.~Shen, X.~G.~Wu, Y.~Ma and S.~J.~Brodsky,
  Phys.\ Lett.\ B {\bf 770}, 494 (2017).

\bibitem{Brodsky:1997jk}
 S.~J.~Brodsky and P.~Huet,
 Phys.\ Lett.\ B {\bf 417}, 145 (1998).

 \bibitem{Costa:2012cb}
 M.~S.~Costa, V.~Goncalves and J.~Penedones,
 JHEP {\bf 1212}, 091 (2012).

 \bibitem{Gromov:2015vua}
 N.~Gromov, F.~Levkovich-Maslyuk and G.~Sizov,
 Phys.\ Rev.\ Lett.\  {\bf 115}, 251601 (2015).

\bibitem{Gromov:2013pga}
N.~Gromov, V.~Kazakov, S.~Leurent and D.~Volin,
Phys.\ Rev.\ Lett.\  {\bf 112}, 011602 (2014).

\bibitem{Gromov:2014caa}
N.~Gromov, V.~Kazakov, S.~Leurent and D.~Volin,
JHEP {\bf 1509}, 187 (2015).

\bibitem{Gromov:2015wca}
N.~Gromov, F.~Levkovich-Maslyuk and G.~Sizov,
JHEP {\bf 1606}, 036 (2016).


\end{thebibliography}
\end{document}